\documentclass[12pt,a4paper]{article}
\pdfoutput=1
\usepackage{lineno}  
\usepackage{graphicx}  
\usepackage{xspace} 
\usepackage{color}
\usepackage{colortbl}
\usepackage{amsmath} 
\usepackage{ifthen} 

\usepackage{booktabs} 
\usepackage[margin=10pt,font={footnotesize},labelfont=bf,labelsep=quad]{caption}

\newboolean{pdflatex}
\setboolean{pdflatex}{true} 

\newboolean{articletitles}
\setboolean{articletitles}{true} 

\newboolean{uprightparticles}
\setboolean{uprightparticles}{false} 

\usepackage{amssymb}
\usepackage{amsfonts}
\usepackage{upgreek} 

\def\lhcb {LHCb\xspace}
\def\ux85 {UX85\xspace}

\def\lhc {LHC\xspace}

\def\rich   {RICH\xspace}

\def\ecal   {ECAL\xspace}

\ifthenelse{\boolean{uprightparticles}}%
{
 
 \def\Pgamma      {\ensuremath{\upgamma}\xspace}

 \def\Ppi         {\ensuremath{\uppi}\xspace}

 \def\Pphi        {\ensuremath{\upphi}\xspace}

 \def\PDelta      {\ensuremath{\Delta}\xspace}                 
 \def\PXi      {\ensuremath{\Xi}\xspace}                 
 \def\PLambda      {\ensuremath{\Lambda}\xspace}                 
 \def\PSigma      {\ensuremath{\Sigma}\xspace}                 
 \def\POmega      {\ensuremath{\Omega}\xspace}                 
 \def\PUpsilon      {\ensuremath{\Upsilon}\xspace}                 
 
 \def\PB      {\ensuremath{\mathrm{B}}\xspace}                 
                  
 \def\PD      {\ensuremath{\mathrm{D}}\xspace}

 \def\PK      {\ensuremath{\mathrm{K}}\xspace}

 \def\PW      {\ensuremath{\mathrm{W}}\xspace}

 \def\Pb      {\ensuremath{\mathrm{b}}\xspace}

 \def\Pi      {\ensuremath{\mathrm{i}}\xspace}

 \def\Pp      {\ensuremath{\mathrm{p}}\xspace}

 \def\Ps      {\ensuremath{\mathrm{s}}\xspace}

}
{
 
 \def\Pgamma      {\ensuremath{\gamma}\xspace}

 \def\Ppi         {\ensuremath{\pi}\xspace}

 \def\Pphi        {\ensuremath{\phi}\xspace}

 \mathchardef\PDelta="7101
 \mathchardef\PXi="7104
 \mathchardef\PLambda="7103
 \mathchardef\PSigma="7106
 \mathchardef\POmega="710A
 \mathchardef\PUpsilon="7107
                  
 \def\PB      {\ensuremath{B}\xspace}                 
                  
 \def\PD      {\ensuremath{D}\xspace}

 \def\PK      {\ensuremath{K}\xspace}

 \def\PW      {\ensuremath{W}\xspace}

 \def\Pb      {\ensuremath{b}\xspace}

 \def\Pi      {\ensuremath{i}\xspace}

 \def\Pp      {\ensuremath{p}\xspace}

 \def\Ps      {\ensuremath{s}\xspace}

}

\def\g      {\ensuremath{\Pgamma}\xspace}

\def\W      {\ensuremath{\PW}\xspace}

\def\squark    {\ensuremath{\Ps}\xspace}

\def\pion  {\ensuremath{\Ppi}\xspace}
\def\piz   {\ensuremath{\pion^0}\xspace}

\def\pim   {\ensuremath{\pion^-}\xspace}

\def\pipm  {\ensuremath{\pion^\pm}\xspace}

\def\kaon  {\ensuremath{\PK}\xspace}
  \def\Kbar  {\kern 0.2em\overline{\kern -0.2em \PK}{}\xspace}

\def\Kz    {\ensuremath{\kaon^0}\xspace}
\def\Kzb   {\ensuremath{\Kbar^0}\xspace}
\def\KzKzb {\ensuremath{\Kz \kern -0.16em \Kzb}\xspace}
\def\Kp    {\ensuremath{\kaon^+}\xspace}
\def\Km    {\ensuremath{\kaon^-}\xspace}

\def\KpKm  {\ensuremath{\Kp \kern -0.16em \Km}\xspace}

\def\Kstarz  {\ensuremath{\kaon^{*0}}\xspace}

  \def\Dbar    {\kern 0.2em\overline{\kern -0.2em \PD}{}\xspace}
\def\D       {\ensuremath{\PD}\xspace}

\def\Dz      {\ensuremath{\D^0}\xspace}
\def\Dzb     {\ensuremath{\Dbar^0}\xspace}
\def\DzDzb   {\ensuremath{\Dz {\kern -0.16em \Dzb}}\xspace}
\def\Dp      {\ensuremath{\D^+}\xspace}
\def\Dm      {\ensuremath{\D^-}\xspace}

\def\DpDm    {\ensuremath{\Dp {\kern -0.16em \Dm}}\xspace}

\def\Dstarpm {\ensuremath{\D^{*\pm}}\xspace}

\def\B       {\ensuremath{\PB}\xspace}
  \def\Bbar    {\kern 0.18em\overline{\kern -0.18em \PB}{}\xspace}

\def\Bz      {\ensuremath{\B^0}\xspace}

\def\Bd      {\ensuremath{\B^0}\xspace}
\def\Bs      {\ensuremath{\B^0_\squark}\xspace}

  \def\Y#1S{\ensuremath{\PUpsilon{(#1S)}}\xspace}

\def\proton      {\ensuremath{\Pp}\xspace}

\def\BF         {{\ensuremath{\cal B}\xspace}}

\def\BR         {\BF}
\newcommand{\decay}[2]{\ensuremath{#1\!\to #2}\xspace}         

\def\to                 {\ensuremath{\rightarrow}\xspace}

\def\BsPhiGam     {\decay{\Bs}{\phi \g}}
\def\BdKstGam     {\decay{\Bd}{\Kstarz \g}}

\def\AT#1     {\ensuremath{A_{\mathrm{T}}^{#1}}\xspace}           

\def\C#1      {\ensuremath{\mathcal{C}_{#1}}\xspace}                       
\def\Cp#1     {\ensuremath{\mathcal{C}_{#1}^{'}}\xspace}                    
\def\Ceff#1   {\ensuremath{\mathcal{C}_{#1}^{\mathrm{(eff)}}}\xspace}        
\def\Cpeff#1  {\ensuremath{\mathcal{C}_{#1}^{'\mathrm{(eff)}}}\xspace}       
\def\Ope#1    {\ensuremath{\mathcal{O}_{#1}}\xspace}                       
\def\Opep#1   {\ensuremath{\mathcal{O}_{#1}^{'}}\xspace}                    

\newcommand{\tev}{\ensuremath{\mathrm{\,Te\kern -0.1em V}}\xspace}
\newcommand{\gev}{\ensuremath{\mathrm{\,Ge\kern -0.1em V}}\xspace}
\newcommand{\mev}{\ensuremath{\mathrm{\,Me\kern -0.1em V}}\xspace}
\newcommand{\kev}{\ensuremath{\mathrm{\,ke\kern -0.1em V}}\xspace}
\newcommand{\ev}{\ensuremath{\mathrm{\,e\kern -0.1em V}}\xspace}
\newcommand{\gevc}{\ensuremath{{\mathrm{\,Ge\kern -0.1em V\!/}c}}\xspace}
\newcommand{\mevc}{\ensuremath{{\mathrm{\,Me\kern -0.1em V\!/}c}}\xspace}
\newcommand{\gevcc}{\ensuremath{{\mathrm{\,Ge\kern -0.1em V\!/}c^2}}\xspace}
\newcommand{\gevgevcccc}{\ensuremath{{\mathrm{\,Ge\kern -0.1em V^2\!/}c^4}}\xspace}
\newcommand{\mevcc}{\ensuremath{{\mathrm{\,Me\kern -0.1em V\!/}c^2}}\xspace}

\def\cm   {\ensuremath{\rm \,cm}\xspace}

\def\mm   {\ensuremath{\rm \,mm}\xspace}

\def\nm   {\ensuremath{\rm \,nm}\xspace}

\def\barn{\ensuremath{\rm \,b}\xspace}

\def\invnb {\ensuremath{\mbox{\,nb}^{-1}}\xspace}

\def\invpb {\ensuremath{\mbox{\,pb}^{-1}}\xspace}

\def\invfb   {\ensuremath{\mbox{\,fb}^{-1}}\xspace}

\def\ns   {\ensuremath{{\rm \,ns}}\xspace}
\def\ps   {\ensuremath{{\rm \,ps}}\xspace}
\def\fs   {\ensuremath{\rm \,fs}\xspace}

\newcommand{\chisq}{\ensuremath{\chi^2}\xspace}

\def\gsim{{~\raise.15em\hbox{$>$}\kern-.85em
          \lower.35em\hbox{$\sim$}~}\xspace}
\def\lsim{{~\raise.15em\hbox{$<$}\kern-.85em
          \lower.35em\hbox{$\sim$}~}\xspace}

\def\pt         {\mbox{$p_{\rm T}$}\xspace}
\def\et         {\mbox{$E_{\rm T}$}\xspace}

\def\dllkpi     {\ensuremath{\mathrm{DLL}_{\kaon\pion}}\xspace}

\def\mrad{\ensuremath{\rm \,mrad}\xspace}
\def\rad{\ensuremath{\rm \,rad}\xspace}

\def\evtgen     {\mbox{\textsc{EvtGen}}\xspace}
\def\pythia     {\mbox{\textsc{Pythia}}\xspace}

\def\geant      {\mbox{\textsc{Geant4}}\xspace}

\def\gauss      {\mbox{\textsc{Gauss}}\xspace}

\def\tell1  {TELL1\xspace}
\def\ukl1   {UKL1\xspace}

\newcommand{\ie}{\mbox{\itshape i.e.}}

\def\photos     {\mbox{\textsc{Photos}}\xspace} 

\ifthenelse{\boolean{uprightparticles}}%
{
 \def\Pphi      {\ensuremath{\upphi}\xspace}
}
{
 \def\Pphi      {\ensuremath{\phi}\xspace}
}

\def\pp    {\ensuremath{\proton\proton}\xspace}

\newcommand{\eq}[1]{Eq.~\ref{equation:#1}}

\newcommand{\tab}[1]{Table~\ref{table:#1}}
\newcommand{\fig}[1]{Fig.~\ref{figure:#1}}

\def\BRBdKstGam {\ensuremath{\BR(\decay{\Bd}{\Kstarz \g})}\xspace}
\def\BRBsPhiGam {\ensuremath{\BR(\decay{\Bs}{\phi \g})}\xspace}

\usepackage{mciteplus}

\begin{document}

\renewcommand{\thefootnote}{\fnsymbol{footnote}}
\setcounter{footnote}{1}


\begin{titlepage}
\pagenumbering{roman}

\vspace*{-1.5cm}
\centerline{\large EUROPEAN ORGANIZATION FOR NUCLEAR RESEARCH (CERN)}
\vspace*{1.5cm}
\hspace*{-0.5cm}
\begin{tabular*}{\linewidth}{lc@{\extracolsep{\fill}}r}
\ifthenelse{\boolean{pdflatex}}
{\vspace*{-2.7cm}\mbox{\!\!\!\includegraphics[width=.14\textwidth]{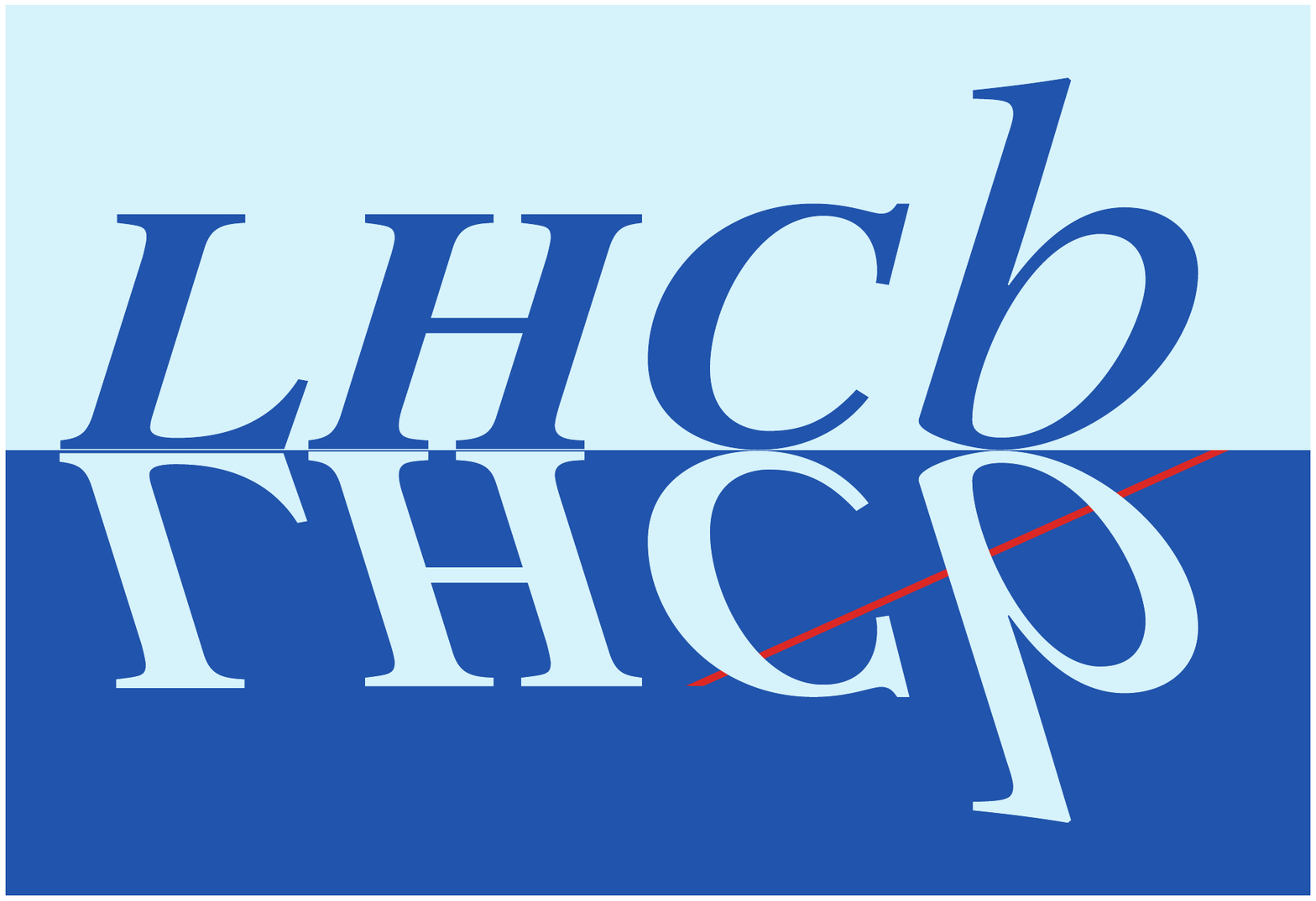}} & &}%
{\vspace*{-1.2cm}\mbox{\!\!\!\includegraphics[width=.12\textwidth]{lhcb-logo.eps}} & &}%
\\
 & & CERN-PH-2012-041 \\  
 & & LHCb-PAPER-2011-042 \\  
 & & \today \\ 
 & & \\
\end{tabular*}

\vspace*{3.0cm}

{\bf\boldmath\huge
\begin{center}
Measurement of the ratio of branching fractions \BRBdKstGam/\BRBsPhiGam
\end{center}
}

\vspace*{2.0cm}

\begin{center}
The LHCb collaboration
\footnote{Authors are listed on the following pages.}
\end{center}

\vspace{\fill}

\begin{abstract}
    \noindent
  The ratio of branching fractions of the radiative \B decays \BdKstGam and \BsPhiGam has been measured using $0.37\,\invfb$ of \pp collisions at a centre of mass energy of $\sqrt{s}=7\,\tev$, collected by the \lhcb experiment.
  The value obtained is 
 \begin{equation}
 \frac{\BRBdKstGam}{\BRBsPhiGam} =  1.12 \pm 0.08^{+0.06}_{-0.04}\phantom{.}^{+0.09}_{-0.08},\nonumber
\end{equation}
where the first uncertainty is statistical, the second systematic and the third is associated to the ratio of fragmentation fractions $f_s/f_d$.
   Using the world average for $\BRBdKstGam = (4.33\pm 0.15)~\times10^{-5}$, the branching fraction \BRBsPhiGam is measured to be $(3.9\pm0.5)\times10^{-5}$, which is the most precise measurement to date.
\end{abstract}

\vspace*{2.0cm}

\begin{center}
Submitted to Physical Review D
\end{center}

\vspace{\fill}

\end{titlepage}


\newpage
\setcounter{page}{2}
\mbox{~}
\newpage

\begin{center}
  {\bf LHCb collaboration}
\end{center}
\begin{flushleft}
R.~Aaij$^{38}$, 
C.~Abellan~Beteta$^{33,n}$, 
B.~Adeva$^{34}$, 
M.~Adinolfi$^{43}$, 
C.~Adrover$^{6}$, 
A.~Affolder$^{49}$, 
Z.~Ajaltouni$^{5}$, 
J.~Albrecht$^{35}$, 
F.~Alessio$^{35}$, 
M.~Alexander$^{48}$
% ...

{\footnotesize \it
% ...
$ ^{5}$Clermont Universit\'{e}, Universit\'{e} Blaise Pascal, CNRS/IN2P3, LPC, Clermont-Ferrand, France\\
$ ^{6}$CPPM, Aix-Marseille Universit\'{e}, CNRS/IN2P3, Marseille, France\\
% ...
$ ^{33}$Universitat de Barcelona, Barcelona, Spain\\
$ ^{34}$Universidad de Santiago de Compostela, Santiago de Compostela, Spain\\
$ ^{35}$European Organization for Nuclear Research (CERN), Geneva, Switzerland\\
% ...
$ ^{38}$Nikhef National Institute for Subatomic Physics, Amsterdam, The Netherlands\\
% ...
$ ^{43}$H.H. Wills Physics Laboratory, University of Bristol, Bristol, United Kingdom\\
% ...
$ ^{48}$School of Physics and Astronomy, University of Glasgow, Glasgow, United Kingdom\\
$ ^{49}$Oliver Lodge Laboratory, University of Liverpool, Liverpool, United Kingdom\\
% ...
$ ^{n}$LIFAELS, La Salle, Universitat Ramon Llull, Barcelona, Spain\\
% ...
}
\bigskip
\end{flushleft}

\cleardoublepage

\renewcommand{\thefootnote}{\arabic{footnote}}
\setcounter{footnote}{0}


\pagestyle{plain} 
\setcounter{page}{1}
\pagenumbering{arabic}


\section{Introduction}

In the Standard Model (SM) the decays \BdKstGam and \BsPhiGam\footnote{Charge conjugated modes are implicitly included throughout the paper.}  proceed at leading order through \decay{b}{s\gamma} one-loop electromagnetic penguin transitions, dominated by a virtual intermediate top quark coupling to a \W boson.
Extensions of the SM predict additional one-loop contributions that can introduce sizeable effects on the dynamics of the  transition~\cite{Descotes:theo-isospin:2011,*Gershon:th-null-tests:2006,*Mahmoudi:th-msugra:2006,*Altmannshofer:2011gn}.

Radiative decays of the $B^0$ meson were first observed by the CLEO collaboration in 1993~\cite{cleo:exp-first-penguins:1993} through the decay mode \decay{\B}{\kaon^{*}\gamma}.
In 2007 the Belle collaboration reported the first observation of the analogous decay in the \Bs sector, $\BsPhiGam$~\cite{belle:exp-bs2phigamma-bs2gammagamma:2007}. 
The current world averages of the branching fractions of  $\BdKstGam$ and $\BsPhiGam$ are $(4.33\pm 0.15)~\times10^{-5}$ and $(5.7^{+2.1}_{-1.8})~\times10^{-5}$, respectively~\cite{hfag:2010,babar:exp-b2kstgamma:2009,*belle:exp-b2kstgamma:2004,*cleo:exp-excl-radiative-decays:1999}.
These results are in agreement with the latest SM theoretical predictions from NNLO calculations using SCET~\cite{Ali:th-b2vgamma-NNLO:2008}, \mbox{$\BR(\BdKstGam) = (4.3\pm 1.4)\times10^{-5}$} and \mbox{$\BR(\BsPhiGam) = (4.3\pm 1.4)\times10^{-5}$}, which suffer from large hadronic uncertainties.
The ratio of experimental branching fractions is measured to be $\BRBdKstGam/\BRBsPhiGam$ = $0.7 \pm 0.3$, in agreement with the prediction of $1.0 \pm 0.2$~\cite{Ali:th-b2vgamma-NNLO:2008}.

This paper presents a measurement of $\BRBdKstGam/\BRBsPhiGam$ using a strategy that ensures the cancellation of most of the systematic uncertainties affecting the measurement of the individual branching fractions.
The measured ratio is used to determine \BRBsPhiGam assuming the world average value of \BRBdKstGam~\cite{hfag:2010}.

\section{The \lhcb detector and dataset}
The LHCb detector~\cite{lhc:lhcb:2008} is a single-arm forward spectrometer
covering the pseudorapidity range $2<\eta <5$, designed for the study of
particles containing $b$ or $c$ quarks. The detector includes a high precision
tracking system consisting of a silicon-strip vertex detector (VELO) surrounding
the $pp$ interaction region, a large-area silicon-strip detector located
upstream of a dipole magnet with a bending power of about 4~Tm, and three stations of
silicon-strip detectors and straw drift-tubes placed downstream. The combined
tracking system has a momentum resolution $\Delta p/p$ that varies from 0.4\% at
5~\gevc to 0.6\% at 100~\gevc, and an impact parameter (IP) resolution of
$20\,\mu$m for tracks with high transverse momentum. Charged hadrons are
identified using two ring-imaging Cherenkov (RICH) detectors. Photon, electron
and hadron candidates are identified by a calorimeter system consisting of
scintillating-pad and pre-shower detectors, an electromagnetic calorimeter and a
hadronic calorimeter. Muons are identified by a muon system composed of
alternating layers of iron and multiwire proportional chambers. 
The trigger
consists of a hardware stage, based on information from the calorimeter and muon
systems, followed by a software stage running on a large farm of commercial
processors which applies a full event reconstruction.

The data used for this analysis correspond to $0.37\,\invfb$ of \pp collisions collected in the first half of 2011 at the \lhc with a centre of mass energy of $\sqrt{s}=7\,\tev$.
\BdKstGam and \BsPhiGam candidates are required to have triggered on the signal photon and vector meson daughters, following a definite trigger path. The hardware level must have been triggered by an \ecal candidate with $\et > 2.5\,\gev$.
In the software trigger, the events are selected when a  track is reconstructed with  IP $\chi^2>16$, and either $\pt>1.7\,\gevc$ when the photon has $\et>2.5\,\gev$ or $\pt>1.2\,\gevc$ when the photon has $\et>4.2\,\gev$.
The selected track must form a \Kstarz or \Pphi candidate when combined with an additional track, and the invariant mass of the combination of the $\Kstarz(\Pphi)$ candidate and the photon candidate is requested to lie within a $1\,\gevcc$ window around the nominal $\Bz(\Bs)$ mass.

Large samples of $\BdKstGam$ and $\BsPhiGam$ Monte Carlo (MC) simulated events~\cite{LHCb-PROC-2011-006} are used to optimize the signal selection and to parametrize the \B meson invariant mass distribution.
The \pp collisions are generated with \pythia 6.4~\cite{Sjostrand:pythia:2006} and decays of hadronic particles are simulated using \evtgen~\cite{Lange:evtgen:2001} in which final state radiation is generated using \photos~\cite{PHOTOS}.
The interaction of the generated particles with the detector and its response are simulated using \geant~\cite{Agostinelli:geant4:2003}.



\section{Event selection}
The selection of both \B decays is designed to ensure the cancellation of systematic uncertainties in the ratio of their efficiencies.
The procedure and requirements are kept as similar as possible:
the $\Bz(\Bs)$ mesons are reconstructed from a selected $\Kstarz(\Pphi$), composed of oppositely charged kaon-pion (kaon-kaon) pairs, combined with a photon.

The two tracks from the vector meson daughters are both required to have  \mbox{$\pt>500\,\mevc$} and to point away from all  \pp interaction vertices by requiring $\text{IP}\,\chi^2 > 25$.
The identification of the kaon and pion tracks is made by applying cuts to the particle identification (PID) provided by the \rich system.
The PID is based on the comparison between two particle hypotheses, and it is represented by the difference in logarithms of the likelihoods (DLL) between the two hypotheses.
Kaons are required to have \mbox{\dllkpi$>5$} and \mbox{DLL$_{\kaon\proton}>2$}, while pions are required to have \mbox{\dllkpi$<0$}.
With these cuts, kaons (pions) coming from the studied channels are identified with a $\sim70\,(83)\,\%$ efficiency for a $\sim3\,(2)\,\%$ pion (kaon) contamination.

Two-track combinations are accepted as $\Kstarz(\Pphi)$ candidates if they form a vertex with $\chi^2<9$ and their invariant mass lies within a $\pm50\,(\pm10)\,\mevcc$ mass window of the nominal $\Kstarz(\Pphi)$ mass.
The resulting vector meson candidate is combined with a photon of  \mbox{$\et>2.6\,\gev$}.
Neutral and charged electromagnetic clusters in the \ecal are separated based on their compatibility with extrapolated tracks~\cite{Deschamps:exp-calo-reco:2003} while photon and \piz deposits 
are identified on the basis of the shape of the electromagnetic shower in the \ecal.
The \B candidate  invariant mass resolution, dominated by the photon contribution, is about $100\,\mevcc$ for the decays presented in this paper.

The \B candidates are required to have an invariant mass within a $\pm800\,\mevcc$ window around the corresponding \B hadron mass, to have $\pt > 3\,\gevc$ and to point to a \pp interaction vertex by requiring $\text{IP}\,\chi^2 < 9$. The distribution of the helicity angle $\theta_{\text{H}}$, defined as the angle between the momentum of either of the daughters of the vector meson ($V$) and the momentum of the \B candidate
 in the rest frame of the vector meson, is expected to follow $\sin^2\theta_{\text{H}}$ for \decay{\B}{V\gamma}, and $\cos^2\theta_{\text{H}}$ for the \decay{\B}{V\piz} background.
Therefore, the helicity structure imposed by the signal decays is exploited to remove \decay{\B}{V\piz} background, in which the neutral pion is misidentified as a photon, by requiring that $|\cos\theta_{\text{H}}|<0.8$.
Background coming from  partially reconstructed \Pb-hadron decays is rejected by requiring vertex isolation:  the $\chi^2$ of the \B vertex must increase by more than half a unit when adding any other track in the event.



\section{Determination of the ratio of branching fractions\label{section:extraction}}

The ratio of the branching fractions is calculated from the number of signal candidates in the \BdKstGam and \BsPhiGam channels,

\begin{equation}
  \frac{\BR(\BdKstGam)}{\BR(\BsPhiGam)} = \frac{N_{\BdKstGam}}{N_{\BsPhiGam}}
                                         \times \frac{\BR(\phi\to K^+K^-)}{\BR(K^{*0}\to K^+\pi^-)}
                                         \times \frac{f_s}{f_d}
                                         \times \frac{\epsilon_{\BsPhiGam}}{\epsilon_{\BdKstGam}},
  \label{equation:BRRatio}
\end{equation}
where $N$ corresponds to the observed number of signal candidates (yield), $\BR(\phi\to K^+K^-)$ and $\BR(K^{*0}\to K^+\pi^-)$ are the visible branching fractions of the vector mesons,  $f_s/f_d$ is the ratio of the $\Bd$ and $\Bs$ hadronization fractions in \pp collisions at $\sqrt{s} = 7\,\tev$, and $\epsilon_{\BsPhiGam}/\epsilon_{\BdKstGam}$ is the ratio of efficiencies for the two decays.
This latter ratio is split into contributions coming from the acceptance ($r_{\text{acc}}$), the reconstruction and selection requirements ($r_{\text{reco}}$), the PID requirements ($r_{\text{PID}}$), and the trigger requirements ($r_{\text{trig}}$)~:
\begin{equation}
  \frac{\epsilon_{\BsPhiGam}}{\epsilon_{\BdKstGam}} = r_{\text{acc}}
                                               \times r_{\text{reco}}
                                               \times r_{\text{PID}}
                                               \times r_{\text{trig}}.
\end{equation}

The PID efficiency ratio is measured from data to be  $r_{\text{PID}}=0.787 \pm 0.010\,\text{(stat)}$, by means of a calibration procedure using pure samples of kaons and pions from \mbox{\decay{\Dstarpm}{\Dz(\Kp\pim)\pipm}} decays selected utilizing purely kinematic criteria.
The other efficiency ratios have been extracted using simulated events.
The acceptance efficiency ratio, $r_{\text{acc}}=1.094 \pm 0.004\,\text{(stat)}$, exceeds unity  because of the correlated acceptance of the kaons due to  the limited phase space  in the \decay{\Pphi}{\Kp\Km}  decay.
These phase-space constraints also cause the \Pphi vertex to have a worse spatial resolution than the \Kstarz vertex. This affects the  \BsPhiGam selection efficiency through the IP \chisq and vertex isolation cuts while the common track cut $\pt>500\,\mevc$ is less efficient on the softer pion from the \Kstarz decay. Both effects almost compensate and the reconstruction and selection efficiency ratio is found to be  $r_{\text{reco}}=0.949 \pm 0.006\,\text{(stat)}$, where the main systematic uncertainties in the numerator and denominator cancel since the kinematic selections are mostly identical for both decays.
The trigger efficiency ratio $r_{\text{trig}}=1.057 \pm 0.008\,\text{(stat)}$ has been computed taking into account the contributions from the different trigger configurations during the data taking period.

\begin{figure}[hptb]
\begin{center}
  \includegraphics[width=0.95\textwidth]{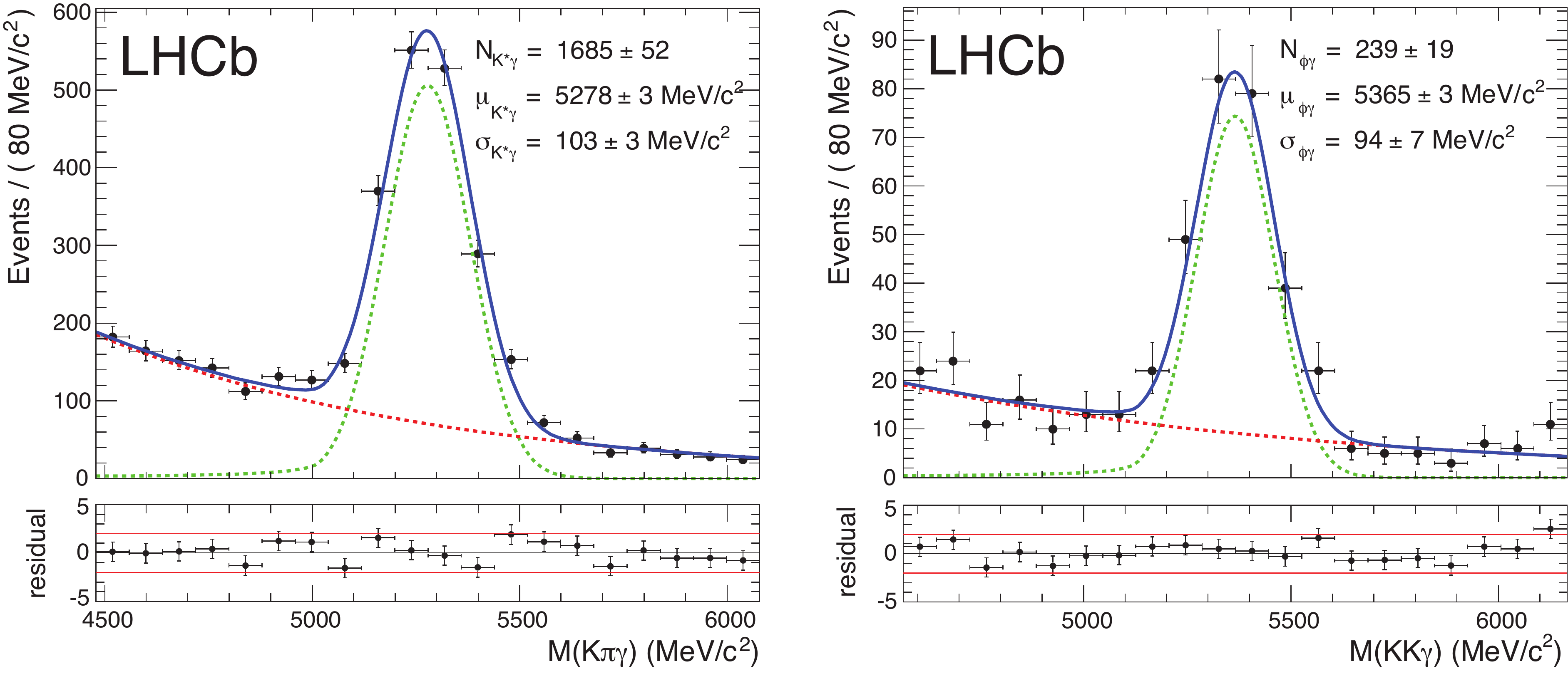}
\caption{Result of the fit for the \BdKstGam (left) and \BsPhiGam (right). The black points represent the data and the fit result is represented as a solid line.
The signal is fitted with a Crystal Ball function (light dashed line) and the background is described as an exponential (dark dashed line).
Below each invariant mass plot, the Poisson \chisq residuals~\cite{Baker:chi2:1984} are shown.\label{figure:fits}}
\end{center}
\end{figure}

The yields of the two channels are extracted from a simultaneous unbinned maximum likelihood fit to the invariant mass distributions of the data.
Signals are described using a Crystal Ball function~\cite{Skwarnicki:cb:1986}, with the tail parameters fixed to their  values extracted from MC simulation and the mass difference between the $\Bd$ and $\Bs$ signals fixed~\cite{pdg:2010}.
The width of the signal peak is left as a free parameter.
Combinatorial background is parametrized by an exponential function with a different decay constant for each channel.
The results of the fit are shown in \fig{fits}.
The number of events obtained for $\BdKstGam$ and $\BsPhiGam$ are $1685\pm52$ and $239\pm19$, with a signal over background ratio of $S/B=3.1\pm0.4$ and $3.7\pm1.3$ in a $\pm3\sigma$ window, respectively.

Several potential sources of peaking background have been studied:
\decay{\B^{0}_{(s)}}{\Kp\pim\piz} and \decay{\Bs}{\Kp\Km\piz}, where the two photons from the \piz can be merged into a single cluster and misidentified as a single photon, \decay{\Lambda_b^0}{\Lambda^{*0}(\kaon p)\,\gamma}, where the proton can be misidentified as a pion or a kaon, and the irreducible \decay{\Bs}{\Kstarz\gamma}.
Their invariant mass distributions and selection efficiencies have been evaluated from simulated events and the number of predicted background events is determined and subtracted from the signal yield.

\B decays in which one of the decay products has not been reconstructed, such as \mbox{\decay{B}{(\Kstarz\piz)X}},  tend to accumulate towards lower values in the invariant mass distribution but can contaminate the signal peak.
However, their contributions have not been included in the fit, and the correction to the fitted signal yield has been quantified by means of a statistical study.
The mass distribution of the partially reconstructed \B decays is first extracted from a sample of simulated events and the corresponding shape has been added to the fit with a free amplitude.
The fit is then repeated many times varying the shape parameters and the  amplitude of the partially reconstructed component within their uncertainties.
The correction to be applied to the signal yield and its uncertainty at a $95\%$ confidence level are determined from the obtained distribution of the signal yield variation.

The effects of the cross-feed between the two channels, \ie~$\BdKstGam$ signal misidentified as $\BsPhiGam$ and vice-versa, as well as the presence of multiple \B candidates per event, have also been computed using simulation.
The statistical uncertainty due to finite MC sample size is taken as the uncertainty in these corrections.

Table \ref{table:background} summarizes all the corrections applied to the fitted signal yields, as well as the corresponding uncertainties, for each source of background.
\begin{table}[hb*]
\begin{center}
\caption{Correction factors and corresponding uncertainties affecting the signal yields, in percent, induced by  peaking backgrounds,  partially reconstructed backgrounds, signal cross-feed and multiple candidates. The total  uncertainty is obtained by summing the individual contributions in quadrature.}\label{table:background}
 \begin{tabular}{ccccccc}

  \toprule
                                         &  \multicolumn{2}{c}{\BdKstGam } &  \multicolumn{2}{c}{\BsPhiGam } &  \multicolumn{2}{c}{   Ratio  }\\ 
 \cmidrule(r{0.2em}l{0.2em}){2-3} 
 \cmidrule(r{0.2em}l{0.2em}){4-5}
  \cmidrule(r{0.2em}l{0.2em}){6-7}
  Contribution                                                           & Corr. &  Error & Corr. &  Error &  Corr. &  Error \\
  \midrule
    \decay{\Bd}{K^+\pi^-\pi^0}                   &   $-1.3$ &    $\pm0.4$       &   ---          & $<0.1$  &   $-1.3$ &    $\pm0.4$\\
    \decay{\Bs}{K^+\pi^-\pi^0}                   &   $-0.5$ &    $\pm0.5$       &   ---          & $<0.1$   &   $-0.5$ &    $\pm0.5$\\
    \decay{\Bs}{K^+K^-\pi^0}                    &       ---            &    $<0.1$     &   $-1.3$  & $\pm 1.3$ &   $+1.3$  & $\pm 1.3$ \\
  \decay{\Lambda_b^0}{\Lambda^{*0}\gamma}                  &    $ -0.7$          &   $\pm 0.2$    &    $-0.3$   &     $\pm 0.2$  & $-0.4$   &     $\pm 0.3$             \\
    \decay{\Bs}{\Kstarz\gamma}             &   $-0.8$  &     $\pm 0.4$      &   ---              &  ---   &   $-0.8$  &     $\pm 0.4$\\
  \midrule
    Partially reconstructed \B                &   $+0.04$               &    ${}^{+3.1}_{-0.2}$      &   $+4.5$               &    ${}^{+1.3}_{-2.9}$&   $-4.5$               &    ${}^{+4.2}_{-1.3}$ \\
  \midrule
   $\phi\gamma/K^{*0}\gamma$ cross-feed                             &   $-0.4$  &     $\pm0.2$      &   ---              & $<0.1$  &   $-0.4$  &     $\pm0.2$ \\
   Multiple candidates                &    $-0.5$   &     $\pm 0.2$  &    $-0.3$   &     $\pm 0.2$         &    $-0.2$   &     $\pm 0.3$    \\
  \midrule
  \bf  Total                                            & \bf   $-4.2$     & \bf    ${}^{+3.2}_{-0.9}$    &   \bf $+2.6$    & \bf${}^{+1.9}_{-3.2}$ &   \bf $-6.8$    & \bf${}^{+4.5}_{-2.0}$ \\
  \bottomrule
  \end{tabular}
\end{center}
\end{table}

The ratio of branching fractions from \eq{BRRatio} is calculated using the fitted yields  of the signal corrected for the backgrounds, the values of the visible branching fractions~\cite{pdg:2010}, the LHCb measurement of $f_s/f_d$~\cite{Aaij:fsfdhadronic:2011,lhcb:fsfd-paper:2011}, and the values of the efficiency ratios described above.
The result is

\begin{equation*}
  \frac{\BR(\BdKstGam)}{\BR(\BsPhiGam)}=1.12\pm0.08\mathrm{(stat)}. \label{equation:resultStat}
\end{equation*}



\section{Systematic uncertainties}

The limited size of the MC sample used in the calculation of $r_{\text{acc}}$, $r_{\text{reco}}$, and $r_{\text{trig}}$ induces a systematic uncertainty in the ratio of branching fractions.
In addition, $r_{\text{acc}}$ is affected by uncertainties in the hadron reconstruction efficiency, arising from differences in the interaction of pions and kaons with the detector and the uncertainties in the description of the material of the detector.
Differences in the mass window size of the vector mesons, combined with small differences in the position of the $\Kstarz(\Pphi)$ mass peaks between data and MC, produce a systematic uncertainty in $r_{\text{reco}}$ which has been evaluated by moving the centre of the mass window to the value found in data.
The reliability of the simulation to describe  the $\text{IP}\,\chi^2$ of the tracks and the \B vertex isolation has been propagated into an uncertainty for  $r_{\text{reco}}$.
For this, the MC sample has been reweighted to reproduce the background-subtracted distributions from data, obtained by applying  the {\it sPlot} technique~\cite{Pivk:stat-splots:2005} to separate signal and background components, using the invariant mass of the $B$ candidate as the discriminant variable.
No further systematic errors are associated with the use of MC simulation, since kinematic properties of the decays are known to be well modelled.
Systematic uncertainties associated with the photon are negligible due to the fact that its reconstruction in both decays is identical.

The systematic uncertainty associated with the PID calibration method has been evaluated using MC simulation.
The statistical error due to the size of the kaon and pion calibration samples has also been propagated to $r_{\text{PID}}$.

The systematic effect introduced by applying a \B mass window cut of $\pm800\,\mevcc$ has been evaluated by repeating the fit procedure with a tighter \B mass window reduced to $\pm 600\,\mevcc$.

\tab{summary-syst} summarizes all sources of systematic uncertainty, including the background contributions detailed in \tab{background}.
The uncertainty on the ratio of efficiency-corrected yields is obtained by combining the individual sources in quadrature. 
The uncertainty on the ratio $f_s/f_d$  is given as a separate source of uncertainty.

\begin{table}[htb!]
\center
\caption{\label{table:summary-syst}Summary of contributions to the relative systematic uncertainty on the ratio of branching fractions.
Note that  $f_s/f_d$ is quoted as a separate systematic uncertainty.}
\begin{tabular}{lc}
   \toprule
  Source                                                   & Uncertainty (\%)         \\ \midrule
  Acceptance ($r_{\text{acc}}$)                            & $\pm0.3$             \\
  Selection ($r_{\text{reco}}$)                       & $\pm1.4 $            \\
  PID efficiencies  ($r_{\text{PID}}$)                     & $\pm2.7 $            \\
  Trigger ($r_{\text{trig}}$)                     & $\pm0.8 $            \\
  \B mass window                                           & $\pm0.9$             \\
  Background                                               & ${}^{+4.5}_{-2.0}$ \\
  Visible fraction of  vector mesons               & $\pm1.0$             \\ \midrule
  Quadratic sum of above                      & ${}^{+5.4}_{-3.3}$ \\ \midrule  
  $f_s/f_d$                                                & ${}^{+7.9}_{-7.5}$ \\ 
  \bottomrule
\end{tabular} 
\end{table}

Besides $f_s/f_d$, the dominant source of systematic uncertainty is the imperfect modelling of the backgrounds due to partially reconstructed $B$ decays.
This specific uncertainty is expected to be reduced when more data are available.



\section{Results and conclusions}

In $0.37\,\invfb$ of \pp collisions at a centre of mass energy of $\sqrt{s}=7$\tev the ratio  of branching fractions of  $\BdKstGam$ and $\BsPhiGam$ decays has been measured to be

\begin{equation*}
 \frac{\BRBdKstGam}{\BRBsPhiGam} =   1.12 \pm 0.08 \mathrm{(stat)} \phantom{.}^{+0.06}_{-0.04} \mathrm{(syst)} \phantom{.}^{+0.09}_{-0.08} (f_s/f_d)
\end{equation*}
in good agreement with the theoretical prediction of $1.0 \pm 0.2$~\cite{Ali:th-b2vgamma-NNLO:2008}.

Using $\BR(\BdKstGam) = (4.33\pm 0.15)~\times10^{-5}$~\cite{hfag:2010}, one obtains
\begin{equation*}
  \BRBsPhiGam = (3.9 \pm 0.5 )\times 10^{-5}
\end{equation*}
(statistical and systematic errors combined), which agrees with the previous experimental value.
This is the most precise measurement of the $\BsPhiGam$ branching fraction to date.

\section*{Acknowledgements}

\noindent We express our gratitude to our colleagues in the CERN accelerator
departments for the excellent performance of the LHC. We thank the
technical and administrative staff at CERN and at the LHCb institutes,
and acknowledge support from the National Agencies: CAPES, CNPq,
FAPERJ and FINEP (Brazil); CERN; NSFC (China); CNRS/IN2P3 (France);
BMBF, DFG, HGF and MPG (Germany); SFI (Ireland); INFN (Italy); FOM and
NWO (The Netherlands); SCSR (Poland); ANCS (Romania); MinES of Russia and
Rosatom (Russia); MICINN, XuntaGal and GENCAT (Spain); SNSF and SER
(Switzerland); NAS Ukraine (Ukraine); STFC (United Kingdom); NSF
(USA). We also acknowledge the support received from the ERC under FP7
and the Region Auvergne.

\bibliographystyle{LHCb}
\bibliography{references}

\end{document}